# Temporal-offset dual-comb vibrometer with picometer axial precision

A. Iwasaki,[1] D. Nishikawa,[1] M. Okano,[1,2] S. Tateno,[1] K. Yamanoi,[1] Y. Nozaki,[1] And S. Watanabe[1,*]

[1]*Faculty of Science and Technology, Keio University, 3-14-1 Hiyoshi, Kohoku-ku, Yokohama, Kanagawa 223-8522, Japan*

[2]*National Defense Academy, 1-10-20 Hashirimizu, Yokosuka-shi, Kanagawa 239-8686, Japan*

*\*Corresponding author: watanabe@phys.keio.ac.jp*

We demonstrate a dual-comb vibrometer where the pulses of one frequency-comb are split into pulse pairs. We introduce a delay between the two pulses of each pulse pair in front of the sample, and after the corresponding two consecutive reflections at the vibrating sample surface, the initially introduced delay is cancelled by a modified Sagnac geometry. The remaining phase difference between the two pulses corresponds to the change in the axial position of the surface during the two consecutive reflections. The Sagnac geometry reduces the effect of phase jitter since both pulses propagate through nearly the same optical path (in opposite directions), and spurious signals are eliminated by time gating. We determine the amplitude of a surface vibration on a surface-acoustic-wave device with an axial precision of 4 pm. This technique enables highly accurate determination of extremely small displacements.

1. **Introduction**

  A quantitative evaluation of extremely small mechanical vibrations of surfaces of solids is important for fundamental physics as well as device applications. For example, the mechanism of interconversion between spin and mechanical angular momentum by surface acoustic waves (SAWs) is a novel route to generate spin currents in nonmagnetic metals [1-4] and opens possibilities for the research area of spin-mechatronics [5]. In order to estimate the conversion efficiency of this mechanism, the evaluation of the surface vibration amplitude is crucial. The evaluation of the vibration amplitude is also useful for the design of SAW filters and microelectromechanical resonators with better efficiencies [6, 7].

  Amplitudes of extremely small surface vibrations are usually determined by optical interferometry. For example, homodyne interferometry enables us to measure absolute vibration amplitudes down to picometers (the phase information can also be obtained) [6, 8]. A sensitivity on the order of femtometers can be achieved [9] by using a heterodyne interferometer [10, 11]. Although the precision of most interferometric experiments based on continuous-wave (cw) lasers is very high, the cw-laser-based approach limits the measurable range, or the so-called ambiguity range, to less than the wavelength of the laser [12]. In addition, spurious reflections can cause additional interferometric signals, which is often troublesome since this degrades the accuracy of the optical-phase measurement [13].

  Multiwavelength interferometry based on two femtosecond-frequency-combs [14] is a novel approach to determine the distance to objects with a large ambiguity range and high precision, which is sometimes referred to as dual-comb ranging. By using two pulsed lasers, time-of-flight distance measurements are possible, which leads to a dramatical extension of the ambiguity range up to kilometers [15, 16]. Such time-of-flight measurements also allow us to easily distinguish between signals from the sample and spurious reflections, which improves the accuracy of the distance measurement [15]. Note that in dual-comb ranging, the optical phase of a pulse reflected from the sample is compared with that of a reference pulse, and this measurement scheme can also be used to perform vibration-amplitude measurements with a large ambiguity range [17]. However, due to the phase jitter of the laser, the precision of such dual-comb vibrometers (which is on the order of sub-nanometers) [17] is still insufficient to probe the amplitude of a SAW vibration, since the latter is on the order of sub-angstroms.

In this article, we present a dual-comb vibrometer with picometer axial precision, which enables us to accurately measure SAW-vibration amplitudes. To achieve this, we first divided one of the frequency combs into two beams by a polarizing beam splitter, and thus each pulse of the original beam was split into a pulse pair with a tunable delay. Due to the delay, the two pulses of a pulse pair were reflected at the vibrating sample at different times corresponding to different axial positions (e.g., maximum and minimum) of the sample surface. Therefore, the difference between the optical phases of the two pulses can be used to determine the SAW amplitude. In our setup, the polarizing beam splitter and the sample are part of a modified Sagnac geometry [18, 19], and thus the two pulses traveled through the same optical components (but in reverse order) before their phase difference was determined by dual-comb spectroscopy (DCS). Note that this geometry dramatically decreases the phase jitter between the two beams. In addition, because of the time-of-flight detection scheme used in DCS, the signal reflected from the sample can be easily separated from other signals, which provides an accurate result free from spurious reflections. We succeeded in determining the absolute vibration amplitude of a SAW with a precision of 4 pm. The SAW propagation was clearly monitored by changing the phase of the SAW with respect to the timing of the laser pulse. This accurate optical measurement technique allows us to determine extremely tiny vibration amplitudes of surfaces, which is useful for investigations of fundamental physics and device design.

**2. Experimental setup**

A schematic of the experimental setup is shown in Fig. 1(a). On the left lower side, two frequency-comb light sources are shown; the two optical frequency combs, which are referred to as the local comb (L-comb) and the signal comb (S-comb), allow us to determine small optical phase shifts of the S-comb pulses that are reflected from the sample. In this work, the phase shift was induced by an extremely tiny surface displacement due to a SAW.

*2.1. Modified Sagnac geometry and sample structure*

The modified Sagnac geometry was used to realize a time-domain analysis of the S-comb. The power and polarization of the S-comb were controlled by a polarizer (POL) and half-wave plate (HWP) located immediately after the exit port of the S-comb light source. After passing through a beam splitter (BS), the S-comb entered the Sagnac geometry unit [Fig. 1(a); inside the green dashed border line]. A polarizing beam splitter (PBS1) was used to divide the S-comb into P-polarized and S-polarized beams, and hereafter we refer to these beams as Beam 1 and Beam 2, respectively. Beam 1 passed through PBS1 and a quarter-wave plate (QWP1), then was reflected by a mirror, and by passing through QWP1 again, Beam 1 was converted to S-polarized light. As a result, Beam 1 was reflected by PBS1 and reached the sample by passing through QWP2 and an objective lens (OL). Meanwhile, Beam 2 was reflected by PBS1 and traveled through a polarization-maintaining (PM) fiber-based polarization rotation unit [Fig. 1(a); inside the gray dashed border line].

The polarization rotation unit used in this work consists of a PM fiber (PM1017-C, Yangtze Optical Fibre and Cable) with a length of about 100 m and a fused fiber polarization combiner/splitter (PFC1550A, Thorlabs), which has three ports: the blue port for both polarization directions, the red port for light polarized along the fast axis, and the white port for light polarized along the slow axis. Beam 2 entered the PM fiber through a fiber collimator (CL) where the polarization of Beam 2 is along the slow axis of the PM fiber. We connected the end of the PM fiber to the blue port of the fiber polarization splitter, and Beam 1 was directed to the white port. To adjust the total fiber length, an additional PM fiber (indicated as a black line) with a length of about 88 cm was connected to the white port. We further connected the exit of the additional PM fiber to the red port in such a way that the slow-axis output couples into the fast axis [Fig. 1(a); "connector part"]. As a result, Beam 2 traveled back to the PM fiber with a polarization along the fast axis, i.e., its polarization was rotated by the polarization unit. After the rotation, Beam 2 passed through PBS1, QWP2, and OL in front of the sample.

Because of the longer traveling distance of Beam 2, the reflection of the pulse of Beam 2 at the sample surface is delayed by $\Delta t$ with respect to the reflection of the pulse of Beam 1. From the total fiber length of the polarization rotation unit and its refractive index ($\approx$1.44)[20], we estimated a $\Delta t$ of about 1.02 µs. We placed the CL on a mechanical delay stage to manually tune $\Delta t$.

The inset on the right upper side of Fig. 1(a) shows a microscopic image of the SAW device. The SAW device consists of patterned titanium and gold layers on a 128° Y-cut 0.5-mm-thick LiNbO$_3$ single crystal substrate fabricated using photolithography and electron-beam evaporation. The thicknesses of the titanium and gold layers were ~3 nm and ~70 nm, respectively, and interdigitated electrodes with a period of ≈21.3 μm were formed on the substrate. We applied a sinusoidal voltage with a frequency of $f_{SAW}$ ≈ 180 MHz to the electrodes to excite Rayleigh-type SAWs on the surface of the LiNbO$_3$ substrate. In this device, the SAW propagation direction is perpendicular to the X-axis of the crystal, and the phase velocity of the SAW is $v_{SAW}$ ≈ 3.67 km/s [21]. The SAW propagates through the rectangular-shaped titanium/gold pad with a size of 400 μm × 360 μm, resulting in a sinusoidal displacement of the surface along the z-direction with a vibration frequency of $f_{SAW}$. The wavelength of the SAW was $\lambda_{SAW}$ = $v_{SAW}$/$f_{SAW}$ ≈ 20 μm. Beams 1 and 2 were focused on the titanium/gold pad by the OL, and since the spot diameter of about 3 μm is sufficiently small with respect to $\lambda_{SAW}$, the local surface vibration at the spot can be probed.

Figure 1(b) illustrates the experimental condition at two different time instants. At times $t = t_1$ and $t = t_1+\Delta t$, the wave packets of Beams 1 and 2, respectively, are reflected from the sample surface, while the surface at the probe spot continuously moves along the z-direction because of the SAW propagation on the LiNbO$_3$ substrate. The difference between the z-positions of the surface at $t_1+\Delta t$ and $t_1$ is defined as displacement $D$. The value of $D$ reaches its maximum if $t_1$ corresponds to a time when the z-position of the surface is at its minimum, and $t_1+\Delta t$ corresponds to a time when the z-position of the surface is at its maximum. This situation is realized if the following three conditions are satisfied: (i) The repetition period of the S-comb is perfectly synchronized with the period of the SAW vibration (or an integral multiple of it), (ii) the interval Δt is equal to half of the period of the SAW vibration (or an odd multiple of the half of the period), and (iii) the reflection of Beam 1 occurs when the z-position of the sample surface is at its minimum. Under these conditions, the displacement $D$ is equal to twice of the SAW amplitude $A_0$, i.e., $D=2A_0$. The strategy used in our experiment to fulfill the above three conditions, is described in Section 2.3.

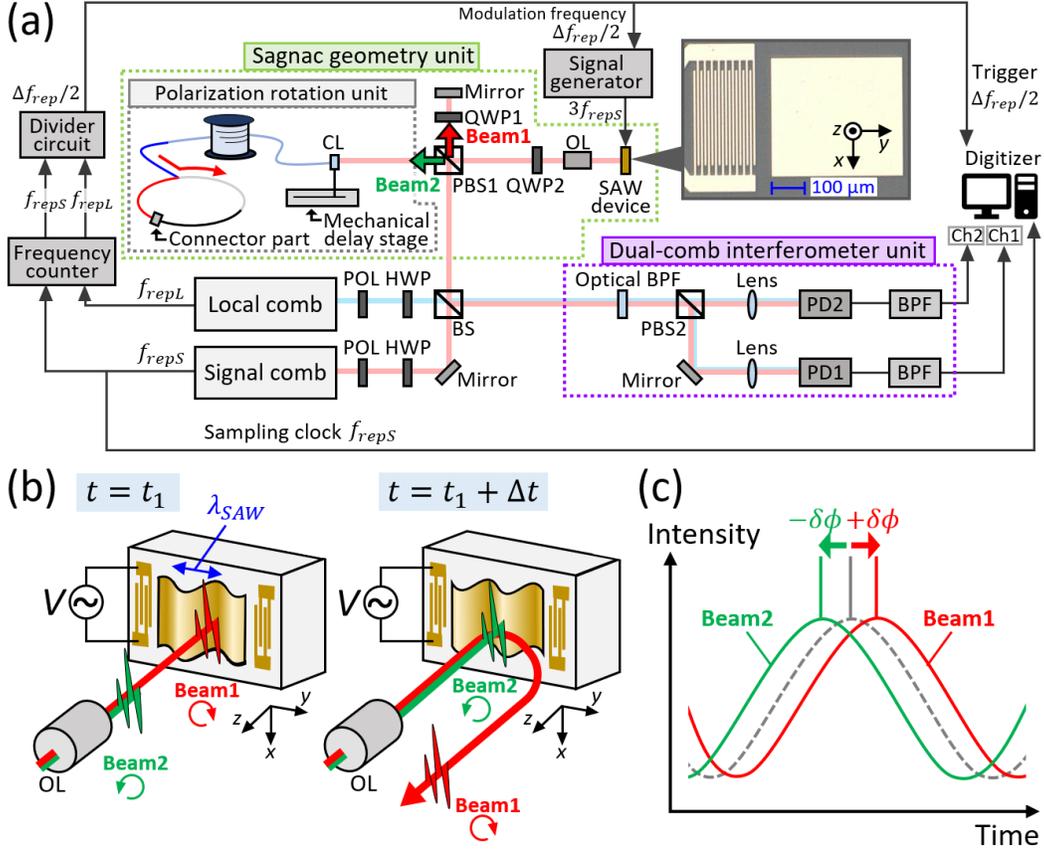

Fig. 1. (a) Experimental setup. BS: non-polarizing beam splitter, PBS1 and PBS2: polarizing beam splitters, POL: polarizer, HWP: half-wave plate, QWP1 and QWP2: quarter-wave plates, OL: objective lens, CL: fiber collimator, PM fiber: polarization-maintaining fiber, BPF: bandpass filter, PD1 and PD2: photodetectors. (b) Experimental condition at times $t=t_1$ and $t=t_1+\Delta t$. The red and green circular arrows indicate the helicities of the circular polarization of Beam 1 and Beam 2, respectively. (c) Temporal profiles of the optical wave at a certain frequency ν (red: Beam 1 and green: Beam 2) at the exit of the Sagnac geometry unit.

When Beams 1 and 2 reached the sample, they were circularly polarized with opposite helicities. Therefore, after reflection at the sample, the two beams traveled through the opposite paths owing to the polarization management [19] as follows: After reflection at the sample, each beam passed through QWP2. Since Beam 1 became P-polarized, it passed through PBS1, traveled through the polarization rotation unit, and returned to PBS1 as an S-polarized beam. Then, Beam 1 was reflected at PBS1 to the exit of the Sagnac geometry unit. Since Beam 2 became S-polarized, it was reflected at PBS1, and passed two times through QWP1 due to the reflection at the mirror behind QWP1. Therefore, it was finally transmitted through PBS1 and left the Sagnac geometry unit as a P-polarized beam.

Now we briefly consider the signal obtained in such a system. If the sample is static ($D=0$), the optical phases of the two beams are almost equal at the exit of the Sagnac geometry unit, because the two beams travel the same optical path in opposite directions. On the other hand, when the sample exhibits a vibration ($D\neq0$), the optical phases of the two beams differ by $2\delta\phi$ as illustrated in Fig. 1(c). The relation between $D$ and $\delta\phi$ is expressed as follows:

$$2\delta\phi = \frac{2D \cdot n_{air}}{\lambda} \times 2\pi \cdots (1),$$

where $\lambda$ is the wavelength of the considered frequency component of the S-comb and $n_{air}$ is the refractive index of air. If the phase difference $\delta\phi$ is precisely measured by DCS, the displacement $D$ can be obtained from Eq. (1).

*2.2. Dual-comb spectroscopy*

To perform the DCS measurement, we first combined the L-comb beam and Beams 1 and 2

exiting from the Sagnac geometry unit by using the BS. The power and polarization of the L-comb beam were controlled by a POL and a HWP as shown in Fig. 1 (a). After combining them, they passed through an optical bandpass filter (BPF) to reduce the bandwidth to such a level that an aliasing effect [14] is avoided when mapping the signal to the radio frequency (RF) domain in the DCS measurement [the dual-comb interferometer unit is shown by the purple dashed border in Fig. 1(a)]. Two interference signals were measured: that between Beam1 and the L-comb and that between Beam 2 and the L-comb. To separately obtain the two interference signals in the RF domain by two photodiodes (PD1 and PD2), we split the combined beam by PBS2. Finally, the electric signals of PD1 and PD2 were recorded using a digitizer in a personal computer with channel 1 (Ch1) and channel 2 (Ch2).

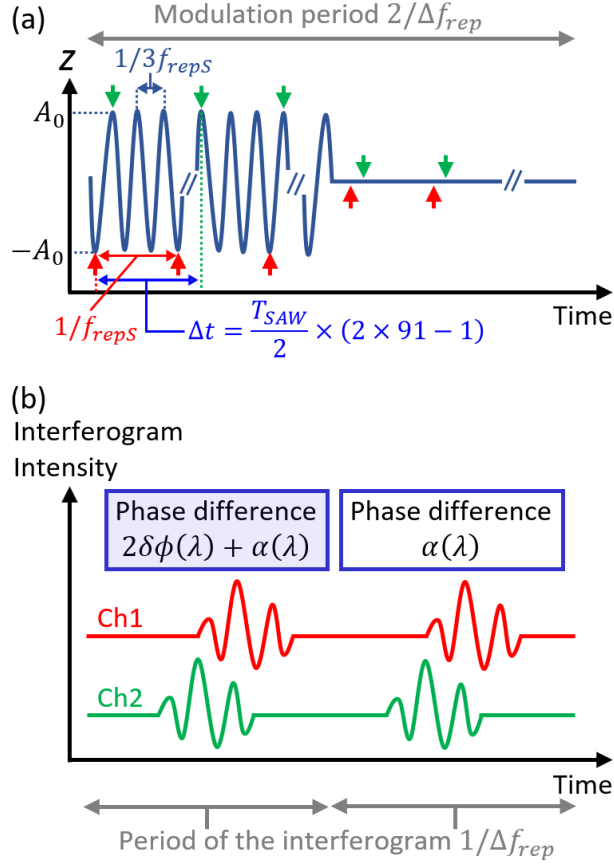

Fig. 2. (a) Schematic of the vibration of the sample surface. Red and green arrows indicate the times when Beam 1 and Beam 2 are reflected at the sample surface, respectively. (b) General relation between the dual-comb interference signals obtained by PD1 and PD2 as a function of the measurement time.

In this paper, three kinds of experiments were conducted by changing the voltage amplitude, the phase, and the sample position (see Section 5 for details). These experiments were conducted on different days and thus had slightly different experimental conditions such as different repetition frequencies of the S-comb, $f_{repS}$, different repetition-frequency offsets between the two frequency combs, $\Delta f_{rep}$ ($\equiv f_{repS} - f_{repL}$), and different ratios of $f_{repS}$ to $\Delta f_{rep}$. The experimental conditions were $f_{repS} \sim 61.531$ MHz, $\Delta f_{rep} \sim 300$ Hz, and $f_{repS}/\Delta f_{rep} \sim 200000$. The ratio $f_{repS}/\Delta f_{rep}$ was chosen to satisfy the coherent averaging condition [22]. The method of the frequency stabilization and the calibration procedure are described in Ref. [23].

*2.3. Frequency synchronization*

To determine the actual SAW amplitude $A_0$ from the measured displacement $D$, it is crucial to synchronize $f_{SAW}$ and $f_{repS}$ as mentioned in Section 2.1. We performed the following procedures to achieve this synchronization: Firstly, we set the frequency of the voltage applied to the SAW device to exactly three times the repetition frequency of the S-comb ($f_{SAW} = 3f_{repS} \sim$ 184.6 MHz). Secondly, the interval $\Delta t$ was optimized by changing the optical path length of the polarization rotation unit by moving the mechanical delay stage. Finally, we tuned the phase of the voltage applied to the SAW device to a value where Beam 1 is reflected from the sample at a time when the *z*-position of the sample surface is at its minimum.

The vibration of the sample surface after the synchronization procedure is shown schematically in Fig. 2(a). We consider the *z*-position of the sample surface at the *y*-position of the beam spot. Under ideal conditions, Beam 1 is reflected at times when the *z*-position of the sample surface is at its minimum ($z = -A_0$). Because we consider $f_{SAW} = 3f_{repS}$, all pulses of Beam 1 are reflected from the sample surface at the same *z*-position. As explained in Section 2.1, we also need to tune $\Delta t$ to a value that fulfills the following equation:

$$\Delta t = \frac{T_{SAW}}{2} \times (2k - 1) \cdots (2),$$

where $T_{SAW}$ is the period of the SAW vibration defined as $T_{SAW} = 1/f_{SAW}$, and $k$ is an integer. If Eq. (2) is fulfilled, Beam 2 is always reflected from the sample surface at times when the *z*-position of the sample surface is at its maximum ($z=A_0$). The integer $k$ is determined by the total optical path length of the polarization rotation unit, and it is around 91 in our experiment. We modulated the sinusoidal voltage applied to the SAW device by a square wave with a modulation frequency of $\Delta f_{rep}/2$. Therefore, as shown on the right-hand side of Fig. 2(a), the vibration of the sample surface is stopped during the second half of the $2/\Delta f_{rep}$ interval (see the gray arrow at the top of the figure).

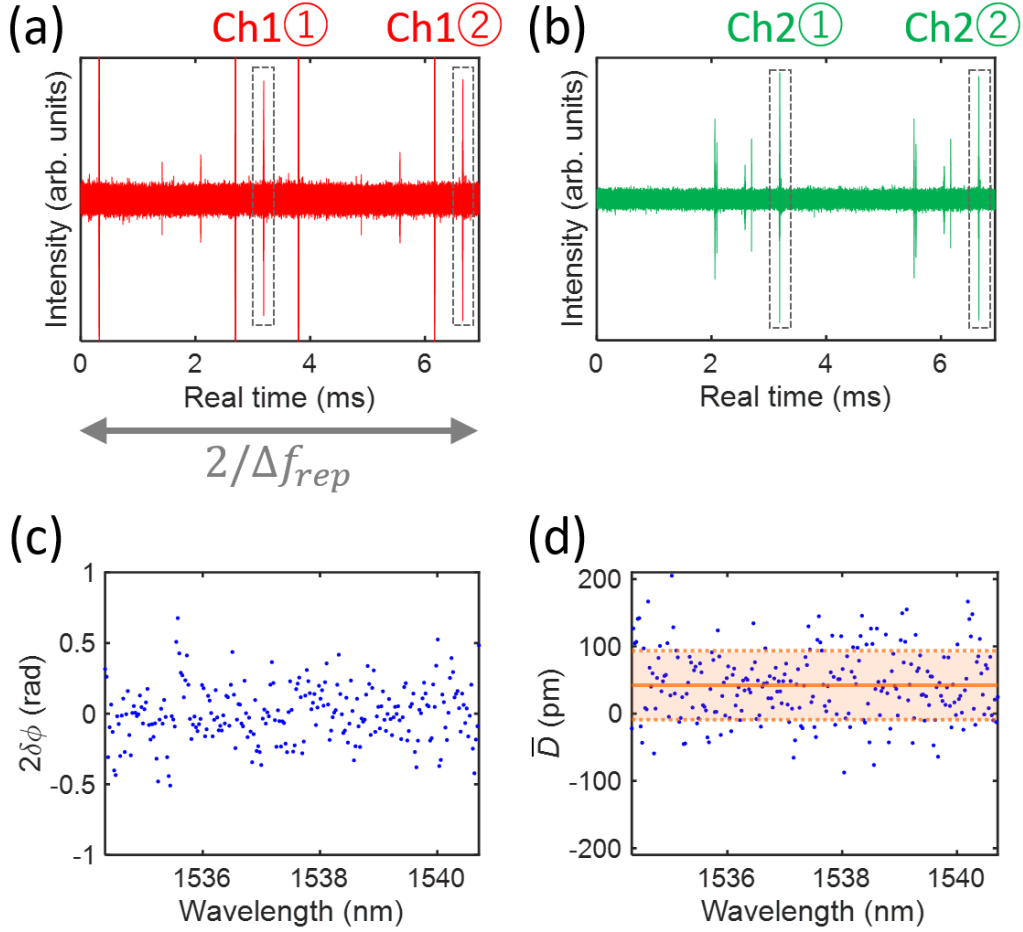

Fig. 3. (a), (b): Measured interferogram data recorded on (a) channel 1 and (b) channel 2 during a single $2/\Delta f_{rep}$ interval when the amplitude of the voltage applied to the SAW device was 0.5 V. The signals originating from the reflection at the sample are indicated by the boxes with dashed lines. (c) Wavelength dependence of the optical phase difference $2\delta\varphi(\lambda)$ associated with the sample vibration obtained by analyzing the optical phases of the signals in the four time-gated interferogram regions in (a) and (b). (d) Wavelength dependence of the average sample displacement $\bar{D}$. The solid and dashed lines represent the mean value of $\bar{D}(\lambda)$, $D_0$, and that including the standard deviation of $\bar{D}(\lambda)$ (defined a $\sigma_{\bar{D}_\lambda} = \sqrt{N} \times \sigma_{D_0}$), $D_0 \pm \sigma_{\bar{D}_\lambda}$.

In our setup, the interferogram originating from the interference between Beam 1 (Beam 2) and the L-comb is recorded on channel 1 (channel 2), and the result is schematically shown by the red (green) curve in Fig. 2(b). Because the update frequency of the interferograms is $\Delta f_{rep}$ and the pulse modulation frequency of the voltage applied to the SAW device is $\Delta f_{rep}/2$, the measured time-domain profiles of each interferogram are repeated with a period of $2/\Delta f_{rep}$. During the first half of the $2/\Delta f_{rep}$ interval, the time-domain data contain the optical phases of Beams 1 and 2 reflected from the vibrating surface, $\varphi_{11}$ and $\varphi_{21}$, respectively. During the second half of the $2/\Delta f_{rep}$ interval, they contain the optical phases of Beams 1 and 2 in the case of no vibration, $\varphi_{12}$ and $\varphi_{22}$, respectively. The phase difference $\alpha = \varphi_{12}-\varphi_{22}$ is a residual phase difference between Beams 1 and 2, which even exists when the vibration is stopped. On the other hand, the phase difference $\varphi_{11}-\varphi_{21}$ is a sum of $\alpha$ and the phase difference $2\delta\varphi$ induced by the SAW vibration [as shown in Fig. 1(c)]. The sample displacement $D$ is determined using Eq. (1) and $2\delta\varphi = (\varphi_{11}-\varphi_{21})-(\varphi_{12}-\varphi_{22})$. The advantage of the present method is the direct comparison of the optical phases of the two pulses of a pulse pair, which are reflected from the sample surface at different times but have propagated through nearly the same optical path (with the same optical components) when they leave the Sagnac geometry unit. Therefore, this setup minimizes the effect of phase jitter. The resulting improved precision is experimentally shown and discussed in the next section.

**3. Experimental results and discussions**

*3.1. Precision of displacement measurements by temporal-offset DCS*

Figures 3 (a) and (b) show the interferogram data recorded on channel 1 and channel 2 during a single $2/\Delta f_{rep}$ interval. In addition to the data originating from the reflection at the sample (the four signals inside the boxes with gray dashed lines), we observed several spurious signals, which originate from reflections at other optical components in the experimental setup. We can easily eliminate these spurious signals by time gating with an appropriate filter window, ensuring an accurate determination of the displacement. For each of the four signals, we used a 0.06-ms-wide cosine-tapered window (or Tukey window) in real time, which corresponds to a width of 0.30 ns in terms of the effective time. Then, we calculated the Fourier transforms of the four time-gated interferograms to obtain the wavelength-dependent phases $\varphi_{11}(\lambda)$, $\varphi_{12}(\lambda)$, $\varphi_{21}(\lambda)$, and $\varphi_{22}(\lambda)$.

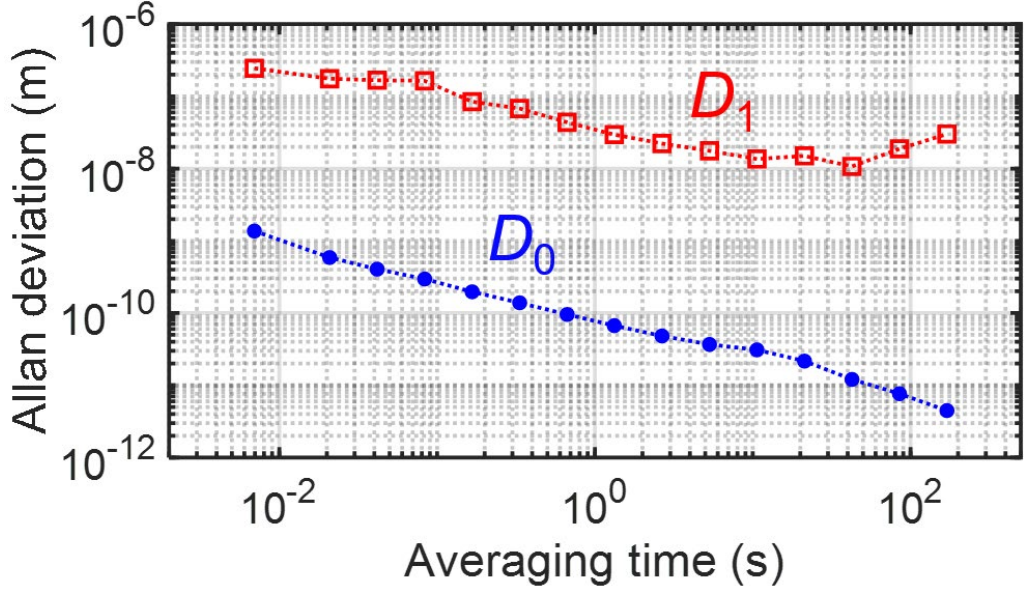

Fig. 4. Averaging-time dependence of the Allan deviation of the distance parameter $D_0$ (blue dots) and that of $D_1$ (red squares).

Figure 3 (c) shows the wavelength dependence of $2\delta\varphi(\lambda)$ [= $\varphi_{11}-\varphi_{21}-\varphi_{12}+\varphi_{22}$] in the range 1534.33 nm ≤ $\lambda$ ≤ 1540.71 nm, where the signal intensity is sufficiently large. The wavelength resolution determined by the time-gating window width is 0.026 nm, resulting in $N$=247 data points in this wavelength range. By using Eq. (1), we then converted the measured $2\delta\varphi(\lambda)$ values to $D(\lambda)$. In this experiment, we recorded 49,000 intervals of length $2/\Delta f_{rep}$ (measurement time less than 3 minutes) and calculated the average of $D(\lambda)$, $\overline{D}(\lambda)$, and its standard deviation of the mean, $\sigma_{\overline{D}(\lambda)}$. Because the value of $\overline{D}$ is in principle independent of the wavelength, we calculated the final result $D_0$ by averaging $\overline{D}(\lambda)$ over the whole wavelength range. Figure 3 (d) shows the wavelength dependence of $\overline{D}(\lambda)$ and $D_0$ is shown by the solid orange line. The standard deviation of the mean of $D_0$, defined as $\sigma_{D_0}$, was calculated according to the law of propagation of uncertainty using the 247 values of $\sigma_{\overline{D}(\lambda)}$. Standard deviation of the 247 values of $\overline{D}(\lambda)$ (defined a $\sigma_{\overline{D}_\lambda} = \sqrt{N} \times \sigma_{D_0}$) is also shown as the dashed lines. We determined that the displacement of the SAW at 0.5 V is $(D_0 \pm \sigma_{D_0}) = (42.3 \pm 3.6)$ pm. Because the time of reflection of Beam 1 coincided with the time when $z = -A_0$, the absolute value of the SAW amplitude is $(A_0 \pm \sigma_{A_0}) = (21.2 \pm 1.8)$ pm.

To discuss the higher precision of the proposed temporal-offset-based multiheterodyne technique, we compare the precisions of the displacement obtained by two different analyses methods using the same experimental data as in Figs. 3 (a) and (b). The first analysis method was used to derive $D_0$ in a similar manner as discussed in the previous paragraph using the pulse pair. Furthermore, we used the Allan deviation of $D_0$ as an indicator of the degree of precision as follows: After the calculation of $D(\lambda)$ of the $i$-th data interval ($1 \leq i \leq 49{,}000$), we calculated the value of $D_0$ for each data interval, $D_0^i$, by averaging $D(\lambda)$ of the $i$-th interval over the whole wavelength range. Then, we calculated the fully overlapping Allan variance of $D_0$ from $D_0^i$ according to Eq. (10) in Ref. [24] as a function of the averaging time. The Allan deviation was calculated by taking the square root of the fully overlapping Allan variance. The second analysis method was used to derive a displacement of the surface by comparing the optical phases of the pulse of Beam1 that were obtained in the cases with and without the SAW vibration. We denote this displacement by $D_1$, which is the displacement that can be obtained by comparing the optical phases of the interferograms with and without the SAW vibration using a single channel. To derive $D_1$, we determined the difference between the phases of Beam1 with and without the applied voltage [$\delta\varphi_1(\lambda)=\varphi_{11}(\lambda)-\varphi_{12}(\lambda)$], converted $\delta\varphi_1(\lambda)$ to $D_1(\lambda)$ for each data interval using Eq. (1), and calculated $D_1^i$ by averaging $D_1(\lambda)$ of the $i$-th data interval over the whole wavelength region. Then, we calculated the Allan deviation of $D_1$ from $D_1^i$ as a function of the averaging time.

Figure 4 shows the averaging-time dependence of the Allan deviation of $D_0$ (blue closed circles) and that of $D_1$ (red open squares). We observe a remarkable improvement of the precision when the distance is estimated by using two beams with a temporal offset of $\Delta t$ instead of a single beam. At short averaging times (< 40 s), the Allan deviations of both $D_0$ and $D_1$ decrease as a function of time, while the Allan deviation of $D_0$ is about two to three orders of magnitude smaller than that of $D_1$. The improved precision in $D_0$ was achieved owing to the comparison of the phases of two pulses generated from the same optical wave packet of the S-comb and the fact that both pulses had propagated through nearly the same optical path when we measured their phases by DCS. This reduces the effect of the phase jitter between Beams 1 and 2 of the S-comb beams. In addition, the two pulses of the pulse pair pass through the same optical components in the experimental setup, which suppresses the effect of subtle fluctuations of the positions of individual optical components. Moreover, the effect of the optical phase fluctuations between the two L-comb beams which interferes with Beams 1 and 2 of the S-comb beams, respectively, was reduced because both pulses interfere with the S-comb at nearly the same timing. At longer averaging times (> 40 s), the Allan deviation of $D_1$ starts to increase. This is probably due to a temporal variation of the phase jitter of the laser pulses between the two successive measurements of the interferogram due to a change in the environmental condition, which results in a drift of the measured value of $D_1$. On the other hand, the Allan deviation of $D_0$ continuously decreases for longer averaging times, and we can confirm a precision of several picometers at 170 s. The precision can become even smaller at longer averaging times; we achieved a 2.5-pm precision using an averaging time of 336 s in a different experimental run. This result also suggests that our experimental scheme is insensitive to temporal variations of the phase jitter because we compare the optical phases of S-comb pulses that have propagated through nearly the same optical path.

*3.2. Characterization of the SAW vibration with picometer precision*

The high precision of the temporal-offset-based dual-comb ranging technique enables us to determine the absolute surface displacement of a vibrating object with an extremely tiny vibration amplitude. In the following, we determine the surface displacement due to a SAW as functions of the amplitude and phase of the applied voltage, and as a function of the *y*-position. Note that the measured displacement $D_0$ depends on the timing of the reflection of Beam 1, since this controls the probed *z*-position of the sample surface; in Fig. 2 (a) we chose a timing corresponding to $z = -A_0$. The timing can be changed by either changing the phase of the applied voltage or by moving the sample along the SAW propagation direction.

In Fig. 5(a), we plot $D_0$ as a function of the phase of the applied voltage for a voltage amplitude of 0.50 V, and we can confirm that $D_0$ exhibits a sinusoidal change. At a phase of 150 degrees (330 degrees), reflection occurs at a time when the z-position of the sample surface is close to its minimum (maximum). The corresponding maximum and minimum values of $D_0$ are 42.3 pm and -41.9 pm, respectively. The absolute values of these extrema are the same within the uncertainty of $\pm 3.6$ pm, proving that an offset-free determination of the displacement is achieved by this technique.

Figure 5(b) shows the results obtained by changing the sample position along the y-direction for 0.50 V. The displacement shows a sinusoidal oscillation along the y-direction, and by fitting the data to the function $D_0 = A_0 \cdot \cos\{2\pi (y - y_0)/\lambda\}$, we obtained $A_0$ = 46.8 pm, $y_0$ = -1.0 μm, and $\lambda$ = 19.2 μm. This result indicates that the SAW is propagating along the y-direction with a wavelength of $\lambda$ that is close to $\lambda_{SAW}$. The observed value of $\lambda$ is slightly smaller than $\lambda_{SAW}$, because we performed the measurement at a position where the thin titanium/gold layers are deposited on the LiNbO$_3$ substrate; it is known that the phase velocity of the SAW is changed by the deposited layers [25, 26]. We found that the amount of displacement at y = 38 μm is slightly smaller than that at y = 18 μm. This may be due to a slight change in the beam spot diameter on the sample surface when we moved the sample, because a larger spot size degrades the spatial resolution and leads to a smaller numerical value of the measured displacement.

Figure 5(c) shows $D_0$ as a function of the applied voltage. A linear dependence of $D_0$ on the applied voltage is clearly observed. The $D_0$ at 0.5 V is smaller than the amplitude determined in Figs. 5(b) and (c). This could be due to an imperfect optimization of the phase of the SAW and/or a degradation of the electric contact.

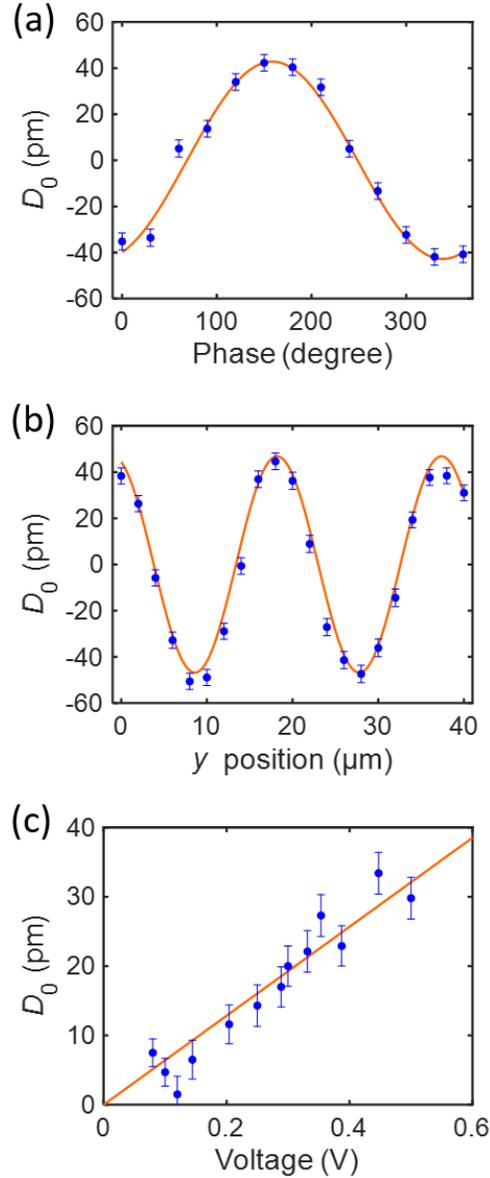

Fig. 5. Characterization of the vibration of the SAW device by the temporal-offset-based multiheterodyne interferometer as a function of (a) the phase of the applied voltage, (b) the sample position, and (c) the applied voltage. For (a) and (b), the amplitude of the applied voltage was set to 0.50 V. The orange lines in (a) and (b) describe the relation between $D_0$ and the phase and that between $D_0$ and the $y$-position, respectively, obtained by a numerical fit of the data to a sinusoidal function. The orange line in (c) describes a linear relation between $D_0$ and the voltage.

## 4. Conclusion

We have demonstrated a new type of dual-comb ranging that is capable of determining the absolute displacement of a vibrating object with picometer axial precision. The comparison of the optical phases of the two pulses of a pulse pair that propagate through nearly the same optical path (with the same optical components) but are reflected at different times from the vibrating sample surface, dramatically improves the axial precision. We determined the amplitude of a surface vibration on a SAW device with an uncertainty of 4 pm. Measurements of the surface displacement as functions of the amplitude and phase of the voltage applied to the SAW device and the sample position clearly prove that a quantitative characterization of surface vibrations can be achieved by this technique. We would like to point out that our technique is capable of sensing a wide range of surface vibrations with different vibration frequencies by changing the distance of one of the arms in the modified Sagnac geometry unit. The achieved precision is the highest among the displacement measurements using dual-comb ranging, and therefore this new type of dual-comb ranging opens an avenue for ultra-precise and accurate length metrology of extremely tiny displacements.

**Funding**

JST CREST (JPMJCR19J4); MEXT Quantum Leap Flagship Program (MEXT Q-LEAP) (JPMXS0118067246).

**Acknowledgements**

S. W. would like to thank Prof. O. B. Wright and Prof. O. Matsuda at Hokkaido University for discussions on the modified Sagnac geometry. A part of this work was conducted at the AIST

Nano-Processing Facility supported by "Nanotechnology Platform Program" of the Ministry of Education, Culture, Sports, Science and Technology (MEXT), Japan, Grant Number JPMXP09-F-21-AT-0085.